# Uncertainty Analysis of the Adequacy Assessment Model of a Distributed Generation System

Yanfu Li[1], Enrico Zio[1,2]

[1] *Chair on Systems Science and the Energetic challenge, European Foundation for New Energy-Electricite' de France, at Ecole Centrale Paris - Supelec, France*

*yanfu.li@ecp.fr, yanfu.li@supelec.fr, enrico.zio@ecp.fr, enrico.zio@supelec.fr*

[2] *Politecnico di Milano, Italy*

*enrico.zio@polimi.it*

**Abstract**

Due to the inherent aleatory uncertainties in renewable generators, the reliability/adequacy assessments of distributed generation (DG) systems have been particularly focused on the probabilistic modeling of random behaviors, given sufficient informative data. However, another type of uncertainty (epistemic uncertainty) must be accounted for in the modeling, due to incomplete knowledge of the phenomena and imprecise evaluation of the related characteristic parameters. In circumstances of few informative data, this type of uncertainty calls for alternative methods of representation, propagation, analysis and interpretation. In this study, we make a first attempt to identify, model, and jointly propagate aleatory and epistemic uncertainties in the context of DG systems modeling for adequacy assessment. Probability and possibility distributions are used to model the aleatory and epistemic uncertainties, respectively. Evidence theory is used to incorporate the two uncertainties under a single framework. Based on the plausibility and belief functions of evidence theory, the hybrid propagation approach is introduced. A demonstration is given on a DG system adapted from the IEEE 34 nodes distribution test feeder. Compared to the pure probabilistic approach, it is shown that the hybrid propagation is capable of explicitly expressing the imprecision in the knowledge on the DG parameters into the final adequacy values assessed. It also effectively captures the growth of uncertainties with higher DG penetration levels.

1. Introduction

Due to the soaring prices of traditional energy sources and the ever-increasing socio-ecological restraints, the power system is experiencing a radical challenge: the evolution from the conventional 'hierarchical structure' to a 'flat structure'. In the former structure, electricity is generated by a small number of centralized and large-sized power plants (e.g. thermal, hydro and nuclear power plants) and is delivered to the end-users through the long-distance transmission network and extensive distribution networks. The latter structure is characterized by the penetration of DG, which enables end-users to install renewable generators (e.g. solar generators and wind turbines) on-site and connect them to the distribution network. This renders the end-users an active player in the production of electricity to satisfy their own demands and even sell it back to the distribution network.

From the perspectives of distribution network operators (DNOs), the major difficulty in the stable management of the emerging DG structure comes from the inherent uncertainties in the operation of renewable generators (Cai et al. 2009, Soroudi and Ehsan 2011). In general, DNOs aim at providing adequate electricity supply to reduce the chance of unsatisfied demand and the consequences of uncertain/risky events in the system. System reliability assessment is performed to reflect the conditions under which the power system is capable of supplying power to the end-users within the specified operating limits. Due to the random nature of renewable generators, uncertainty analysis becomes an unavoidable step in the reliability assessment of the distributed generation (Hegazy et al. 2003, El-Khattam et al. 2006, Atwa et al. 2010).

In the existing literature of DG system reliability assessment, the random behaviors of the renewable generators are typically modeled by two techniques: analytical state enumeration (Billinton and Allan 1996) and Monte Carlo simulation (Azbe and Mihalic 2006, El-Khattam et al. 2006). Most of the existing studies are developed on the assumption that all types of uncertainties in DG can be represented by random variables *X*, described in terms of probability density functions (PDFs), $f(x)$. This type of uncertainty is usually referred to as objective, aleatory, stochastic randomness due to the inherent variability in the system behavior (Apostolakis 1990).

Another type of uncertainty enters the system reliability assessment, due to the incomplete knowledge and information on the system and related phenomena which leads to imprecision in the model representation of the system and in the evaluation of its parameters. This type of uncertainty is often referred to as subjective, epistemic, state-of-knowledge (Apostolakis 1990). In the field of power system research, the epistemic uncertainty has already been considered in the fuzzy power flow analysis (Matos and Gouveia 2008) where the power injections of all loads and generations are regarded as fuzzy variables.

In real-world management of DG systems, e.g. for distribution system asset management (Catrinu and Nordgard, 2011), the DNOs have to confront both aleatory and epistemic uncertainties. However in the DG system reliability assessment studies, the co-existence of aleatory and epistemic uncertainties has not been addressed, except for the very recent work by Soroudi and Ehsan (2011). Moreover, to the knowledge of the authors no previous research has focused on extensively identifying and classifying the uncertainties in DG systems.

Aleatory and epistemic uncertainties may require different mathematical representations and analyses, depending on the information available (Aven and Zio, 2011a,b). When there is limited information to establish probability distributions for the uncertainties in the system model, the possibility distribution is a promising alternative representation of epistemic uncertainties (Baudrit and Dubois, 2006).

For instance, it is common that solar irradiation and wind incidence be modeled by probabilistic distributions, given sufficient historical climate data at the location area of the distribution network; on the contrary, the operation parameters of the renewable generators (e.g. cut-in speed of wind turbine, ambient temperature of solar panel) may be best modeled by possibilistic distributions, for instance because renewable generators are private property of the end-users and it depends on them whether or not to disclose the information of these parameters to DNOs. Even if the end-users were willing to provide this information, it could still be incomplete and inaccurate because the renewable generator manufacturers seldom intend to provide the detailed information about the parameters due to commercial reasons (Rose and Hiskens 2008). Also, in the existing studies these parameters are typically treated as constants in the system model and throughout its life time, although in reality they often vary during the system operation due to the degradation of materials, changes in the operating environments, etc (Giannakoudis et al. 2010).

In the present work, the issues of identifying, classifying, representing and propagating the hybrid (probabilistic and possibilistic) uncertainties in DG systems are systematically addressed within the framework of evidence theory (Shafer, 1976) for processing imprecision and variability.

The paper is organized as follows. In Section 2, a relatively comprehensive distributed generation system model is considered. In Section 3, the related uncertainties are identified and classified. In Section 4, evidence theory and the algorithm for uncertainties propagation are presented. Section 5 provides the case study analyzed. Section 6 concludes the work by discussing findings and limitations.

## 2. Distributed Generation System Model

This Section describes a model for the reliability assessment of a representative distributed generation system. It consists of a number of generation and consumption units. The description is derived from (Li and Zio, 2011). The generation units include renewable generators, e.g. solar generators, wind turbines, and electrical vehicles (EV), and the conventional power source by way of transformers (Figure 1). The transmission lines are often left out of consideration in the reliability assessment studies (Hegazy et al. 2003, Karki et al. 2010). The consumption units can be different types of loads, e.g. residential, commercial, and industrial loads (El-Khattam et al. 2006).

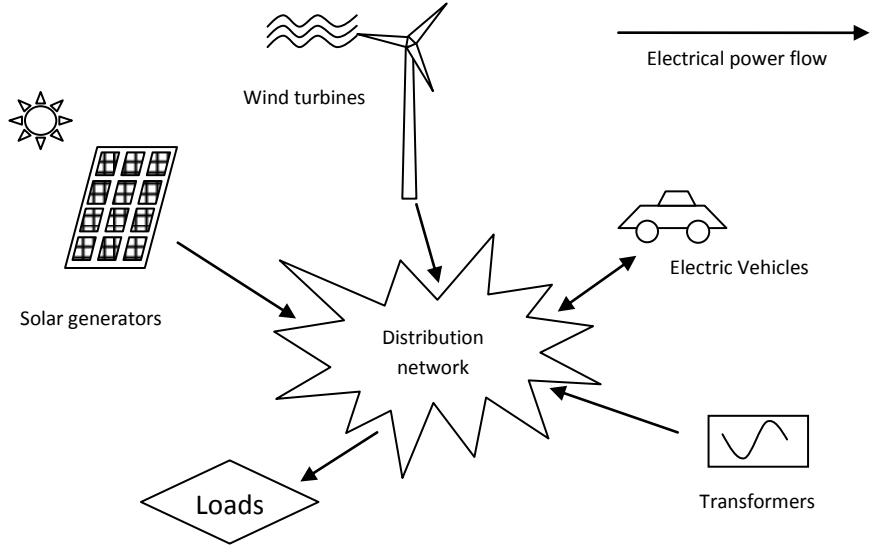

Fig 1. Sketch of the representative distributed generation system

Adequacy/reliability assessment focuses on evaluating the sufficiency of facilities within the system to satisfy the consumer demand (Billinton and Allan, 1996) (i.e. power generation $P_G$ exceeding load power consumption $P_L$):

$$P_A = P_G - P_L \tag{1}$$

Power generation $P_G$ consists of two parts: power from the transmission system, $P_T$ and power from the distributed generators, $P_{DG}$,

$$P_G = P_T + P_{DG} \tag{2}$$

Considering the DG units of Figure 1, this compound power output $P_{DG}$ is:

$$P_{DG} = P_S + P_W + P_{EV} \tag{3}$$

where $P_S = \sum_{i=1}^{m_S} P_i^S$, $P_W = \sum_{i=1}^{m_W} P_i^W$, and $P_{EV} = \sum_{i=1}^{m_{EV}} P_i^{EV}$ are the power outputs from the group of $m_S$ solar generators, $m_W$ wind turbines, and $m_{EV}$ electrical vehicles, respectively, with $P_i^S, P_i^W$, and $P_i^{EV}$ individual power outputs. Note that the value of $P_{EV}$ is negative when the EV group is charging batteries (i.e. consuming power from the network).

In literature, Monte Carlo simulation (MCS) is the mainstream tool for adequacy assessment studies (Billinton et al. 2009, El-Khattam et al. 2006, Hegazy et al. 2003). Three types of MCS techniques have been introduced: sequential MCS (El-Khattam et al. 2006), pseudo-sequential MCS (Leite da Silva et al. 2000), and non-sequential MCS (Veliz et al. 2010). The non-sequential MCS samples the state of all components and combines them to form the system state; it is most efficient, providing comparable accuracy to sequential MCS in shorter execution time (Veliz et al. 2010). In our work, the non-sequential MCS is used.

**2.1 Solar Generator**

In the solar generation group, each photovoltaic (PV) unit is made of a number of solar cells. The model of the *i*th solar generator unit consists of two parts: the solar irradiation function and the power generation function which links the solar irradiation to the power output of the PV solar generator. In literature, the Beta PDF has been used to represent the random behavior of the solar irradiation for each day (Atwa et al. 2010, Zeng et al. 2011):

$$f(s_i) = \begin{cases} \frac{\Gamma(\alpha_i+\beta_i)}{\Gamma(\alpha_i)\Gamma(\beta_i)} \cdot s_i^{(\alpha_i-1)} \cdot (1-s_i)^{\beta_i-1} & for\ 0 \leq s_i \leq 1, \alpha_i \geq 0, \beta_i \geq 0 \\ 0 & otherwise \end{cases} \quad (4)$$

where $s_i \in [0,1]$ is the solar irradiance (measured in kW/m$^2$) received by the *i*th solar generator, $f(s_i)$ is the Beta PDF of $s_i$, $\alpha_i$ and $\beta_i$ are the parameters of the Beta PDF which can be inferred from estimates of the mean and variance values of historical irradiance data (Conti and Raiti 2007). It is noted that if the local distribution network is in a geographical close area, it is typical to assume that $s_i = s, \forall i \in \{1, \dots, m_S\}$.

Once the irradiation distribution is modeled, the output of the *i*th solar generator can be determined by the following power generation function (Mohamed and Koivo 2010):

$$P_i^S = g_S(s_i, \boldsymbol{\theta_i^S}) = N_i \cdot FF_i \cdot V_{y_i} \cdot I_{y_i}$$

$$I_{y_i} = s_i \cdot [I_{SC_i} + k_{c_i}(T_{c_i} - 25)]$$

$$V_{y_i} = V_{oc_i} - k_{v_i} \cdot T_{c_i}$$

$$T_{c_i} = T_{a_i} + s_i \cdot \frac{N_{ot_i} - 20}{0.8}$$

$$FF_i = \frac{V_{MPP_i} \cdot I_{MPP_i}}{V_{oc_i} \cdot I_{sc_i}} \tag{5}$$

where $P_i^S$ is the output power of the $i$th solar generator, $g_S(\cdot)$ is the solar generation function, $\boldsymbol{\theta}_i^S$ is the operation parameter vector of the $i$th solar generator, $k_{v_i}$ is the voltage temperature coefficient V/°C, $k_{c_i}$ is the current temperature coefficient A/°C, $FF_i$ is the fill factor, $I_{sc_i}$ is the short circuit current in A, $V_{oc_i}$ is the open-circuit voltage in V, $I_{MPP_i}$ is the current at maximum power point in A, $V_{MPP_i}$ is the voltage at maximum power point in V, $N_{ot_i}$ is the nominal operating temperature in °C, $T_{c_i}$ is the cell temperature in °C, $T_{a_i}$ is the ambient temperature in °C, $N_i$ is the total number of solar cells in the $i$th solar generator.

## 2.2 Wind Turbine

Similar to the solar, the wind turbine generation model consists of two parts: wind speed modeling and the turbine generation function. The Weibull distribution has been used to model the wind speed randomness (Boyle 2004):

$$f(v_i) = \frac{k_i}{c_i} \left(\frac{v_i}{c_i}\right)^{k_i-1} exp\left[-\left(\frac{v_i}{c_i}\right)^{k_i-1}\right] \tag{6}$$

where $v_i \geq 0$ is the speed of the wind onto the $i$th wind turbine, $k_i$ is the shape index, $c_i$ is the scale index of the $i$th wind turbine, respectively. When $k_i$ equals to 2, the probability density function is called Rayleigh density function. Also in this case, it is typical to assume that $v_i = v, \forall i \in \{1, \ldots, m_W\})$, if the distribution network is located in a geographical close area.

Given the wind speed distribution, the output of the $i$th wind turbine can be modeled by the following function (Zeng et al. 2011):

$$P_i^W = g_W(v_i, \boldsymbol{\theta}_i^W) = \begin{cases} 0 & v_i < v_{ci_i} \text{ or } v_{co_i} \leq v_i \\ P_{r_i} \cdot \frac{(v_i - v_{ci_i})}{(v_{r_i} - v_{ci_i})} & v_{ci_i} \leq v_i < v_{r_i} \\ P_{r_i} & v_{r_i} \leq v_i < v_{co_i} \end{cases} \quad (7)$$

where $\boldsymbol{\theta}_i^W$ is the operation parameter vector of the *i*th wind turbine, $v_{ci_i}$, $v_{co_i}$, $v_{r_i}$, and $P_{r_i}$ are the cut-in wind speed, cut-out wind speed, rated wind speed and rated power output of the *i*th wind turbine, respectively.

## 2.3 Electrical Vehicles

Electrical Vehicles (EVs) can be important elements for distributed generation, with increasing expectation for their positive penetration of the system (Saber and Venayagamoorthy, 2011). The power profile of one individual EV, $P_i^{EV}$ can be negative, zero and positive, because it has a battery storage capable of charging, discharging and holding the power (Clement-Nyns et al. 2011). In our model, a group of $m_{EV}$ EVs is considered distributed on the system. Typically, these are modeled as behaving like a single 'block group' and their power profiles are aggregated as a compound load, source or storage (Clement-Nyns et al. 2011). The physical reasons for grouping EVs into one block are as follows: 1) the battery storage of one individual EV is too small to have influence on the power grid; 2) the majority of the vehicles follow a nearly stable daily usage schedule.

## 2.4 Transformer

The transformer is a stationary device and it is still the major power source in most distributed generation systems. Although the power output from the transformer is often regarded as stable, there are two explicit influential factors that introduce instability into its operation. These factors are the fluctuations of the grid power (Hegazy et al. 2003), and the mechanical degradation/failure/repair of transformer hardware (Ding et al. 2011).

The grid power is represented by a distribution (Hegazy et al. 2003) and the mechanical degradation/failure/repair process is represented by a Markov model (Massim et al. 2006).

**2.5 Load**

In practice, the load values are typically recorded hourly on a specified time horizon (e.g. a year). To model the dynamic behavior of loads, many multi-state probabilistic models have been proposed ranging from a single load-aggregated representation up to more complex individual load modeling (Veliz et al. 2010). Load-aggregated models resort to clustering techniques (Singh and Lago-Gonzales 1989) to reduce the number of load levels, and consider only one geological area pattern; differently, individual load modeling eventually resorts to a multilevel non-aggregate Markov model (Leite da Silva et al. 2000) which considers each hour as one state and includes the changing patterns in different areas. To keep the number of load states limited, we consider the aggregated modeling paradigm.

**3. Identifying and Classifying Uncertainties in Distributed Generation Systems**

**3.1 Uncertainties in Solar Generator Units**

In reminiscence of Section 2.1, the power function of the *i*th solar generator can be written as:

$$P_i^S = g_S(s_i, \boldsymbol{\theta}_i^S) \tag{8}$$

Solar irradiation is typically modeled by a probabilistic distribution (e.g. Beta distribution), because the historical solar irradiation data is often sufficient and accessible to justify such representation (it is measured and recorded) (Atwa et al. 2010, Conti and Raiti, 2007).

The operation parameters $\boldsymbol{\theta}_i^S$ of solar generator unit *i* can be grouped into two categories. One category contains 'coefficients' with values regarded as constant throughout the life time of the solar generator. They are: $k_v$, $k_i$, $I_{sc}$, $V_{oc}$, $I_{MPP}$, $V_{MPP}$, and $N_{ot}$ (the definitions of them are presented in Section 2.1). These parameters are given by the manufacturers. However, due to commercial reasons the manufacturers seldom disclose the detailed information about these parameters (AbdulHadi et al. 2004); they may deliver simplified correlations and models, but the associated uncertainties remain unknown. The other category contains the 'variable parameters' (e.g. ambient temperature $T_a$) which needs to be assessed by the users. Due to privacy issues, the information about some of these parameters can be very limited (e.g. $T_a$ of each household). Consequently, experts' judgments and consumers' behavior knowledge have to be incorporated into the estimation of the operation parameters of the solar generation model: this information is inherently imprecise.

From the above, it seems reasonable to represent solar irradiation as a probabilistic variable and the operation parameters as possibilistic variables. However, this representation is dependent on the information available and it may change from case to case. For instance, if the historical solar irradiation data in a certain area were also insufficient, then the solar irradiation variable may also need to be modeled by possibilistic distributions; on the other hand if the consumers were to provide informative historical records of operation temperatures, then this might suggest the use of probabilistic distributions.

### 3.2 Uncertainties in Wind Turbines

The wind turbines model can have a similar classification of the uncertainties as the solar generators model. In reminiscence of Section 2.2, the power function of the *i*th wind turbine is written as:

$$P_i^W = g_W(v_i, \boldsymbol{\theta}_i^W) \qquad (9)$$

Wind speed is typically modeled by a probabilistic distribution (e.g. Weibull distribution), because the historical wind speed data is often sufficient and accessible to suggest such representation (Billinton et al. 2009, Hong and Pen, 2010).

The operation parameters $\boldsymbol{\theta}_i^W$ of the *i*th wind turbine model can be considered all as 'coefficients'. The coefficients are: $v_{ci_i}$, $v_{co_i}$, $v_{r_i}$, and $P_{r_i}$. These parameters are provided by the manufacturers. But, information about their uncertainties is given with limitations (Rose and Hiskens 2008). Similarly to the treatment of solar generation parameters, we adopt a probabilistic distribution for the wind speed and possibilistic distributions for wind turbine operation parameters.

### 3.3 Uncertainties in Electrical Vehicles

As discussed in Section 2.3, all EVs distributed on the network are treated as a single aggregation with three power output states possible: charging ($P_{EV} < 0$), disconnection ($P_{EV} = 0$), and discharging ($P_{EV} > 0$). Differently from solar and wind generators, EVs power outputs are primarily influenced by the activities of their drivers, who can decide the amount of energy to be exchanged with the grid and the timing/location for the exchange. Due to privacy issues, it might be difficult to gather informative operation data for each EV, so that the estimation of the model parameters relies on expert judgments and knowledge of drivers behavior which is necessarily imprecise. Therefore, the possibilistic distribution is chosen to model the uncertainties in EV power. A similar case is found in Soroudi and Ehsan (2011) where the possibilistic distribution is used to model a general version of renewable generator.

### 3.4 Uncertainties in Transformers and Loads

As anticipated in Section 2.4, there are two types of uncertainties in the operation of transformers: fluctuations of the grid and hardware degradation. In the end, due to the inherent fluctuations in the grid, the power output of the transformer in its working state varies from 80% to 100% of its capacity (Hegazy et al. 2003). Also, we consider that the degradation and failure mechanisms of the transformers have been extensively studied and that there is sufficient information to estimate the parameters of probabilistic distributions assumed to describe them. Finally for the DNO, the real-time load values are usually well monitored by the metering devices installed at the load points and sufficient information can be regarded available to establish a probabilistic representation of the associated uncertainties.

## 3.5 Summary of the Uncertainties in the DG System Model

The following Table 1 summarizes the uncertainties in the DG system model of Section 2.

Table 1. Uncertainties in the DG system model

| Component | Parameter | Source of uncertainty | Type of Information available | Uncertainty representation |
|---|---|---|---|---|
| Solar generator | Solar irradiation | Irradiation variability | Historical data | Probabilistic (e.g. Beta) |
| | Operation parameters | Incomplete knowledge | Experts' judgments, users' experiences | Possibilistic |
| Wind turbine | Wind speed | Speed variability | Historical data | Probabilistic (e.g. Weibull) |
| | Operation parameters | Incomplete knowledge | Experts' judgments, users' experiences | Possibilistic |
| EV aggregation | Power output | Incomplete knowledge, subjective decisions | Experts' judgments, users' experiences | Possibilistic |
| Transformer | Grid power | Power fluctuations | Historical data | Probabilistic |
| | Time to failure | Mechanical degradation/failure date | Historical data | Probabilistic |
| Loads | Load value | Consumption variability | Historical data | Probabilistic |

The overall adequacy assessment model of the DG system can be written as:

$$P_A = f(s_1, \ldots, s_{m_S}, v_1, \ldots, v_{m_W}, P_L, \tilde{P}_{EV}, \tilde{\boldsymbol{\theta}}_1^S, \ldots, \tilde{\boldsymbol{\theta}}_{m_S}^S, \tilde{\boldsymbol{\theta}}_1^W, \ldots, \tilde{\boldsymbol{\theta}}_{m_W}^W) \quad (10)$$

where the possibilistic variables are denoted by the symbol (~). It is observed that the system adequacy output is a function of both aleatory and epistemic uncertain variables and parameters.

## 4. Uncertainty Modeling Methodologies

### 4.1 Probabilistic uncertainty modeling

In the situations that the uncertainty of the variables is mainly due to inherent randomness and there is sufficient information to assign probability distributions and estimate their parameters, probabilistic modeling is due. The model output is represented by a function of $n$ random variables, $Y = f(X_1, \ldots, X_i, \ldots, X_n)$, where $X_i$ denotes the $i$th probabilistic input variable with PDF $p_{X_i}(x)$. Such distribution can be found analytically in simple cases, and by MCS in more realistic settings. In power system studies, the latter is typically used, given the large number of variables involved and their complex relationships, which make analytical models difficult or even impossible to derive (Karki et al. 2010). The operative procedure of MCS calls for a number $m$ of iterations: at each $e$th iteration, an input vector of values $(x_1^e, x_2^e, \ldots, x_n^e)$ is sampled from the PDFs of the input variables and a realization of the output value $y^e$ is computed solving the system model. After $m$ repetitions, an empirical estimate of the distribution of the system output is obtained.

**4.2 Possibilitic uncertainty modeling**

In possibility theory, epistemic uncertainty in the value of a parameter $\tilde{X}$ is modeled by the possibility distribution $\pi_{\tilde{X}}(x)$. For each element $x$ in the set $\Omega$, $\pi_{\tilde{X}}(x)$ represents the degree of possibility that $\tilde{X}$ has value $x$. If there is an element $x_i$ that makes $\pi_{\tilde{X}}(x_i) = 0$, then $x_i$ will be regarded as an impossible outcome. On the other hand, if $\pi_{\tilde{X}}(x_i) = 1$, then $x_i$ will be regarded as a definitely possible outcome, i.e. an unsurprising, normal, usual outcome (Dubois 2006): this is a much weaker statement than the situation when probability equals to 1, which makes the value $x_i$ certain and the value $x_j \neq x_i$ impossible.

Possibility bounds can be defined based on the possibility function. The possibility measure (plausibility) of an event $A$, $Pos(A)$ is defined by:

$$Pos(A) = \sup_{\{x \in A\}} \pi_{\tilde{X}}(x) \tag{11}$$

The necessity measure $Nec(A)$ is defined by:

$$Nec(A) = 1 - Pos(\text{not } A) = 1 - \sup_{\{x \notin A\}} \pi_{\tilde{X}}(x) \tag{12}$$

The possibility measure $Pos$ verifies:

$$\forall A, B \subseteq \mathbb{R} \quad Pos(A \cup B) = \max(Pos(A), Pos(B)) \tag{13}$$

The necessity measure $Nec$ verifies:

$$\forall A, B \subseteq \mathbb{R} \quad Nec(A \cap B) = \min(Pos(A), Pos(B)) \tag{14}$$

The possibility measures can be linked to probabilities in the following manner (Baudrit and Dubois, 2006). Let $\wp(\pi)$ denote a family of probability distributions such that for all events $A$, $Nec(A) \leq Pro(A) \leq Pos(A)$. Then,

$$Pos(A) = \sup Pro(A) \quad \text{and} \quad Nec(A) = \inf Pro(A) \tag{15}$$

where sup and inf are with respect to all probability distributions in $\wp$. Hence, the possibility measure is represented as an upper limit for the probability and the necessity measure is represented as a lower limit.

A typical example of possibility representation is provided below, for illustrative purpose (Baraldi and Zio 2008). Let $\tilde{X}$ be an uncertain parameter which can take values $x \in [1, 4]$, with the most possible values in [2, 3]: the trapezoidal possibility function of Figure 2 can be used to describe the information on the values of $\tilde{X}$ in [1, 4].

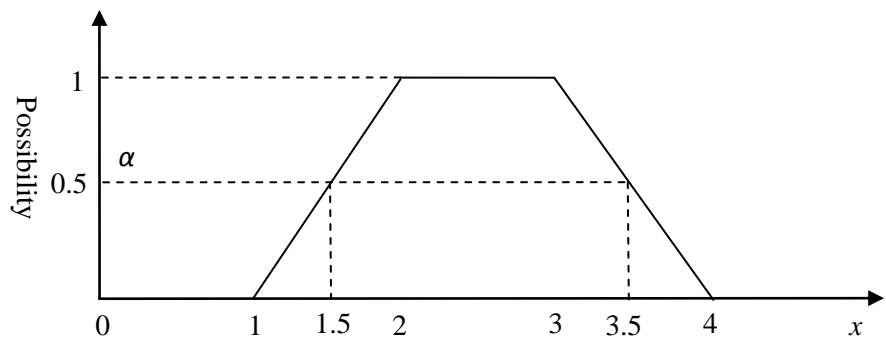

Fig 2. Possibilistic distribution of $\tilde{X}$

**4.2.1 $\alpha - Cut$ method**

The possibilistic output $\tilde{Y}$ of a model of possibilistic inputs $\tilde{X}_i$ is a multivariate function $\tilde{Y} = f(\tilde{X}_1, \tilde{X}_2, \ldots, \tilde{X}_n)$. Given the possibility distributions of the uncertain input variable $\tilde{X}$, it is possible to infer the possibility distribution of $\tilde{Y}$ by means of the $\alpha$-cut method. For a given input variable $\tilde{X}$, we define the $\alpha$-cut of $\tilde{X}$ as:

$$F_\alpha = \{x \in U | \pi_{\tilde{X}}(x) \geq \alpha, 0 \leq \alpha \leq 1\}$$

$$F_\alpha = [\underline{F}_\alpha, \overline{F}_\alpha] \tag{16}$$

where $U$ is the universe of discourse of $\tilde{X}$ (i.e. the range of its possible values), $\underline{F}_\alpha$ and $\overline{F}_\alpha$ are the lower and upper limits of the $\alpha$-cut, respectively. For example, $F_{0.5} = [1.5, 3.5]$ is the set of $x$ values for which the possibility function is greater than or equal to 0.5 (Figure 2): we conclude that if the event $A$ indicates that the parameters lie in the interval [1.5, 3.5], then $0.5 \leq Pro(A) \leq 1$.

Given the $\alpha$-cuts of each uncertain input parameter, the $\alpha$-cut of the output $Y$ can be obtained as:

$$Y_\alpha = [\underline{Y}_\alpha, \overline{Y}_\alpha] \tag{17}$$

$$\underline{Y}_\alpha = \inf f\left(F_\alpha^{\tilde{X}_1}, F_\alpha^{\tilde{X}_2}, \ldots, F_\alpha^{\tilde{X}_n}\right)$$

$$\overline{Y}_\alpha = \sup f\left(F_\alpha^{\tilde{X}_1}, F_\alpha^{\tilde{X}_2}, \ldots, F_\alpha^{\tilde{X}_n}\right)$$

where $F_\alpha^{\tilde{X}_i}$ represents the $\alpha$-cut of the $i$th possibilistic input variable. We note that for each $\alpha$-cut of the output $\tilde{Y}$, the maximum and minimum outputs (upper bound $\overline{Y}_\alpha$, and lower bound $\underline{Y}_\alpha$) are obtained.

## 5. Evidence Theory and Joint Propagation of Probabilistic and Possibilistic Uncertainties

### 5.1 Basic Notions of Evidence Theory

In probability theory, the probability mass (in the discrete case) or probability density (in the continuous case) is assigned to each possible value of a variable, whereas in evidence theory, the variable $X$ takes subsets as its values and probability masses (>0) are assigned to the subsets. Then, a mass distribution $(v_i)_{i=1,\ldots,K}$ can be defined on all the subsets by attaching each mass value $v_i$ to the corresponding subset $A_i$. The mass distribution must satisfy $\sum_{i=1}^{K} v_i = 1$. The portions of mass $v_i$ can be further assigned to specific elements of the subset $A_i$, while elements of $A_i$ may remain without mass due to imprecision and lack of knowledge (Baudrit et al. 2006).

The evidence theory (Shafer, 1976) provides two indicators to quantitatively describe uncertainty with respect to a set $B$: the belief $Bel(B)$ and the plausibility $Pl(B)$ functions; these two qualify the validity of the statement that the values of the variable $X$ (with mass distribution $v(A)$) fall into set $B$. Mathematically, $Bel(B)$ and $Pl(B)$ are defined as:

$$Bel(B) = \sum_{A_i, A_i \subseteq B} v(A_i) \text{ and } Pl(B) = \sum_{A_i, A_i \cap B \neq \emptyset} v(A_i) = 1 - Bel(\bar{B}) \qquad (18)$$

$Bel(B)$ gathers the imprecise evidence that asserts $B$, while $Pl(B)$ gathers the imprecise evidence that does not conflict with $B$. Therefore, the interval $[Bel(B), Pl(B)]$ contains all probability values induced by the mass distribution $v(A)$ on the subset $A$. It is proved that the mass distribution $v$ is the generalization of the probability distribution $p$ and possibility distribution $\pi$ of uncertain discrete variables (the continuous variables have to be discretized) (Baudrit et al. 2006).

**5.2 Algorithm of Joint Propagation of Probabilistic and Possibilistic Uncertainties**

Consider a general power adequacy model $Y = f(X_1, \ldots X_k, \tilde{X}_{k+1}, \ldots, \tilde{X}_n)$ of $n$ uncertain variables $X_i, i = 1, \ldots, n$, ordered in such a way that the first $k$ variables are described by probability distributions $\left(p_{X_1}(x), \ldots, p_{X_k}(x)\right)$, whereas the last $n$-$k$ variables are possibilistic represented by possibility distributions $\left(\pi_{\tilde{X}_{k+1}}(x), \ldots, \pi_{\tilde{X}_n}(x)\right)$. The propagation of the hybrid uncertainty can be performed by MCS combined with the extension principle of fuzzy set theory (Zadeh, 1965) by means of the following two major steps (Baudrit et al. 2006):

1. Repeated Monte Carlo sampling to process the uncertainty in probabilistic variables.
2. Fuzzy interval analysis for treating the uncertainty in possibilistic variables.

The detailed algorithm (Baraldi and Zio 2008, Flage et al. 2010) to calculate the fuzzy random output can be summarized as follows:

---

For $i = 1, 2, \ldots, m$ (the outer loop processing aleatory uncertainty), do:

1. Sample the $i$th realization $(x_1^i, x_2^i, \ldots, x_k^i)$ of the probabilistic variable vector $(X_1, X_2, \ldots, X_k)$.
2. For $\alpha = 0, \Delta\alpha, 2 \cdot \Delta\alpha, \ldots, 1$ (the inner loop processing epistemic uncertainty; $\Delta\alpha$ is the step size, e.g. $\Delta\alpha=0.05$), do:

    2.1 Calculate the corresponding $\alpha$-cuts of possibility distributions $\left(\pi_{\tilde{X}_{k+1}}, \ldots, \pi_{\tilde{X}_n}\right)$ as the intervals of the possibilistic variables $(\tilde{X}_{k+1}, \ldots, \tilde{X}_n)$.

    2.2 Compute the minimal and maximal values of the outputs of the model $f(X_1, \ldots, X_k, \tilde{X}_{k+1}, \ldots, \tilde{X}_n)$, denoted by $\underline{f_\alpha^i}$ and $\overline{f}_\alpha^i$, respectively. In this computation, the probabilistic variables are fixed at the sampled values $(x_1^i, x_2^i, \ldots, x_k^i)$ whereas the possibilistic variables take all values within the ranges of the $\alpha$-cuts of their possibility distributions $\left(\pi_{\tilde{X}_{k+1}}, \ldots, \pi_{\tilde{X}_n}\right)$.

    2.3 Record the extreme values $\underline{f_\alpha^i}$ and $\overline{f}_\alpha^i$ as the lower and upper limits of the $\alpha$-cuts of $f(x_1^i, x_2^i, \ldots, x_k^i, \tilde{X}_{k+1}, \ldots, \tilde{X}_n)$.

    End

3. Cumulate all the lower and upper limits of different $\alpha$-cuts of $f(x_1^i, x_2^i, \ldots, x_k^i, \tilde{X}_{k+1}, \ldots, \tilde{X}_n)$ to establish an approximated possibility distribution (denoted by $\pi_i^f$) of the model output.

End

This procedure results in an ensemble of *m* realizations of the approximated possibility distributions $(\pi_1^f, ..., \pi_m^f)$. For each set *A* in the universe of discourse of all power adequacy values, the following formulas are used to obtain the possibility measure $Pos_i^f(A)$ and the necessity measure $Nec_i^f(A)$, given the possibility distribution $\pi_i^f$:

$$Pos_i^f(A) = \sup_{\{x \in A\}}\{\pi_i^f(x)\}$$

$$Nec_i^f(A) = \inf_{\{x \in A\}}\{1 - \pi_i^f(x)\} \tag{19}$$

These *m* different possibility and necessity measures are then used to obtain the belief $Bel(A)$ and the plausibility $Pl(A)$ of any set *A*, respectively (Baudrit et al. 2006):

$$Pl(A) = \sum_{i=1}^{m} p_i Pos_i^f(A)$$

$$Bel(A) = \sum_{i=1}^{m} p_i Nec_i^f(A) \tag{20}$$

where $p_i$ is the probability of sampling the *i*-th realization $(x_1^i, x_2^i, ..., x_k^i)$ of the random variable vector $(X_1, ..., X_k)$. For each set *A*, this algorithm computes the probability-weighted average of the possibility measures associated with each output fuzzy interval.

## 5.3 Probabilistic Propagation

For pure probabilistic propagation, the possibilistic distributions have to be converted into probability density functions. This conversion can be achieved by various techniques (Flage et al. 2009), e.g. in this paper by simple normalization:

$$p_{X_i}(x) = \frac{\pi_{\tilde{X}_i}(x)}{\int_0^{+\infty} \pi_{\tilde{X}_i}(x) dx} \tag{21}$$

Once the probabilistic distribution for each fuzzy variable is determined, the outer loop of the algorithm in Section 5.2 is performed *m* times, and at each iteration, the vector $(X_1, X_2, ..., X_n)$ is sampled and the corresponding adequacy value is calculated. After the *m* repetitions, the empirical probability distribution of system adequacy is obtained.

## 6. Case Study

The system used as case study is modified from the IEEE 34 node distribution test feeder, and is a radial distribution network downscaled to 230V representing a local residential distribution. In this network, the rated power of the transformer is 5000 kW (Kersting 1991). The modification involves adding a number of renewable generators and distributing them onto the network. To investigate the impacts of different penetration levels of renewable energy, the ratios of renewable energy to conventional energy are set to be 15%, 25% and 35%, respectively. Within the renewable energy, wind, solar, and EV occupy a share of 60%, 30% and 10%, respectively. The DG system infrastructure consists of 3, 5, and 7 identical wind turbines with rated power of 150 kW for the three different penetration levels, respectively; 3, 5, and 7 identical solar generators/arrays (each one containing 1000 solar modules with 75 W rated power), respectively; and 15, 25 and 35 identical EVs with rated power 5 kW, respectively.

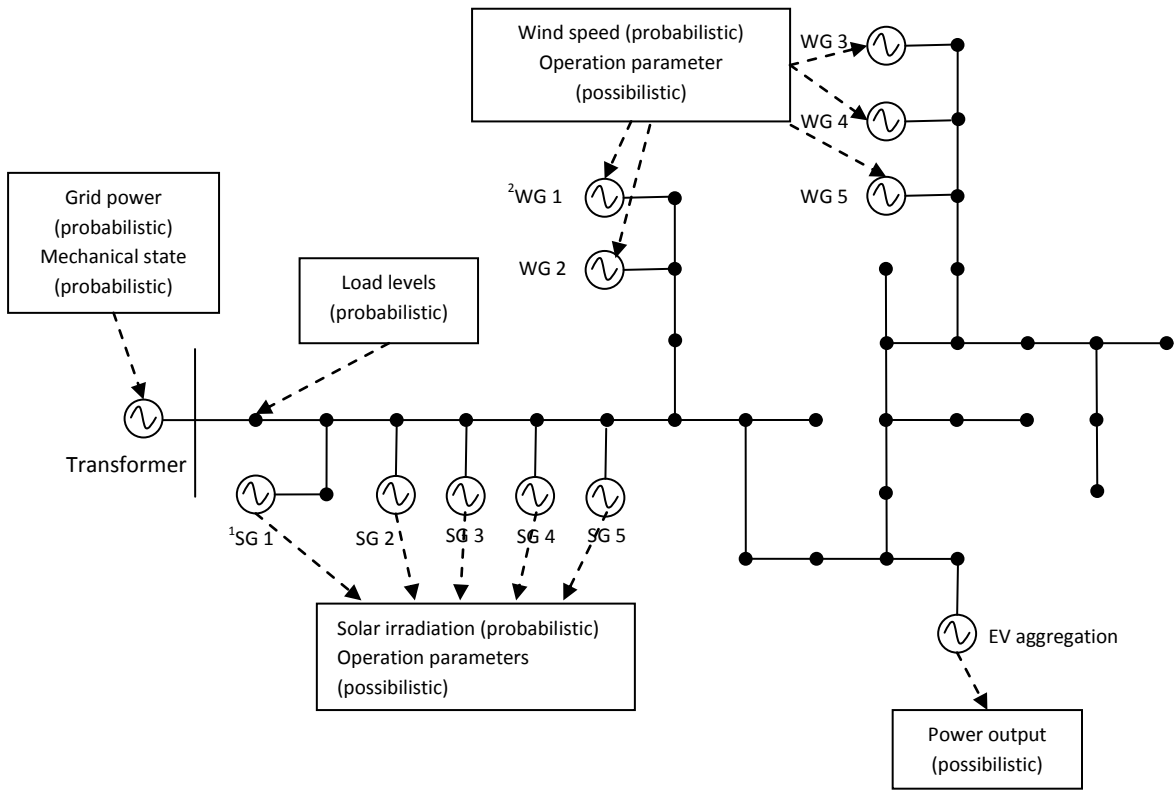

Fig 3. IEEE 34 nodes distribution test feeder modified for distributed generation for 25% renewable penetration level
[1] SG: solar generator, [2] WG: wind generator

## 6.1 Representation of the Uncertainties in the Components of the IEEE 34 DG System Model

The failure mechanism of the transformer is represented by a Markov model with two states: perfectly working and completely failed. The failure and repair rates are 0.0004/yr and 0.013/yr (Roos and Lindah 2004), respectively. No significant uncertainty is assumed to affect the values of these parameters. After solving the Markov model, the steady-state probabilities of the working and failure states are 0.97 and 0.03, respectively. In addition, due to the inherent fluctuation in grid power, the power output of the transformer in its working state is uniformly distributed in the range [0.8, 1.0] of its capacity (Hegazy et al. 2003).

For the group of solar generators, the solar irradiation distribution is the same for all solar generators because it is assumed that the solar distribution system lies in a geographically close area. The parameters of the Beta distribution of solar irradiation have been estimated by fitting the average daily solar irradiation data taken from Mohamed and Koivo (2010). As for the parameters of the solar generation function, their information may typically be incomplete to the DNO (Section 3.1) and a possibilistic distribution is built for each parameter. The summary of the descriptions of the uncertainties of the parameters in the solar generation model is given in Fig 4 and Table 2.

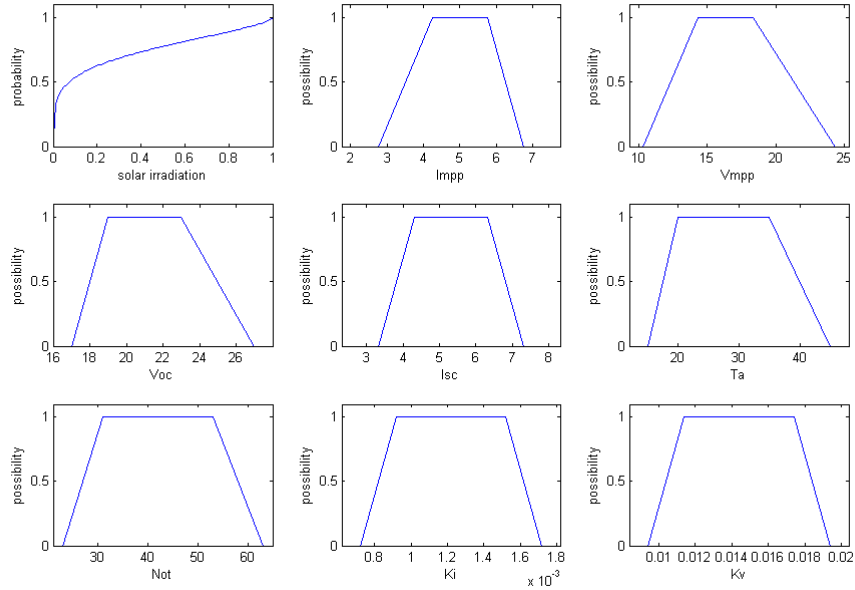

Fig 4. Parameter distributions (probability and possibility) of the solar generator model

Table 2. Parameters of the solar generator distributions

| Possibilistic parameters | Core | Support |
|---|---|---|
| $I_{MPP}$ | [4.56, 4.86] | [4.36, 5.06] |
| $V_{MPP}$ | [16.32, 18.02] | [15.32, 18.32] |
| $V_{oc}$ | [20.98, 21.98] | [19.98, 22.98] |
| $I_{sc}$ | [5.12, 5.42] | [4.82, 5.62] |
| $T_a$ | [29, 30.5] | [27, 32] |
| $N_{ot}$ | [41, 44] | [39, 46] |
| $k_i$ | [0.00112, 0.00132] | [0.00102, 0.00152] |
| $k_v$ | [0.0134, 0.0144] | [0.0124, 0.0164] |
| Probabilistic parameter | $\alpha$ | $\beta$ |
| Solar irradiation | 0.2114 | 0.6454 |

Similar to the solar, the wind speed for the wind turbine group is the same for all members. The parameters of the wind speed distribution (Section 3.2) have been estimated by fitting the average daily wind speed data taken from Mohamed and Koivo (2010). The trapezoidal possibilistic distribution is built for each wind generation parameter (Fig 5 and Table 3).

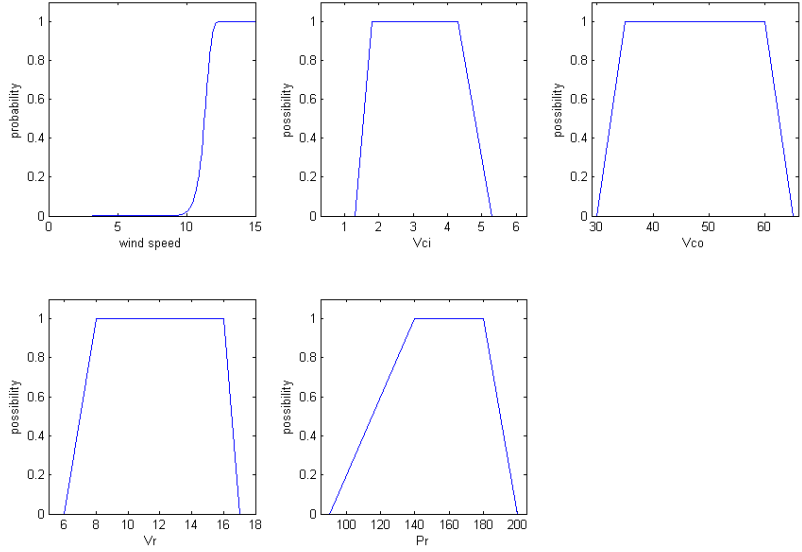

Fig 5. Parameter distributions (probability and possibility) of wind turbine model

Table 3. Parameters of the wind turbine distributions

| Possibilistic parameters | Core | Support |
|---|---|---|
| $v_{ci}$ | [3.2, 3.4] | [3.0, 3.5] |
| $v_{co}$ | [48, 51] | [45, 54] |
| $v_r$ | [11, 11.5] | [10, 12] |
| $P_r$ | [145, 155] | [140, 160] |
| Probabilistic parameter | $\lambda$ | $k$ |
| Wind speed | 18.2304 | 10.4655 |

The power profile of EV aggregation depends on the usage profiles (Section 3.3). The trapezoidal function of core (value sets of possibility equal to 1) [-3, 3], and support (values sets of possibility not equal to zero) [-5, 5] for each EV, is used to model the associated uncertainties.

The hourly peak load curve of the IEEE-RTS is used, with an annual peak load of 4500 kW. This value satisfies the ratio of average peak load to average transformer power output in Hegazy et al. (2003). To perform the non-sequential simulation, the load distribution is divided into ten equally-sized intervals of a histogram for a reasonable trade-off between modeling accuracy and evaluation efficiency (Singh and Lago-Gonzales 1989). The probability for each load interval/state is defined as the ratio of the number of load values in the interval to the total number of load values. The representative value of each interval/state is the average of the lower and upper bounds of the interval.

## 6.2 Results of Uncertainty Propagation in the DG System Model for Adequacy Assessment

After all probability and possibility distributions have been assigned to the model variables and parameters, the uncertainty propagation algorithm of Section 5.2 has been run with $m = 1000$ iterations. At each iteration, the step value of $\alpha$ is set to 0.02 for the calculation of the random fuzzy intervals of output adequacy values. The results obtained are compared against the pure probabilistic approach of uncertainty propagation in (21) and 1000 samples of the joint vector of all parameters in the DG system model are drawn.

Figures 6-8 present the graphical comparisons between the empirical cumulative distribution function (CDF) obtained by the probabilistic propagation approach and the belief and plausibility functions obtained by the joint propagation approach, at different renewable penetration levels. The following observations can be drawn from the comparisons: 1) the CDF of DG adequacy obtained by the pure probabilistic approach lies within the boundaries of belief and plausibility functions obtained by the joint propagation approach; 2) there is an explicit separation between the belief and plausibility functions, in reflection of the total imprecision in the knowledge on the renewable generators parameters; 3) the separation between belief function and plausibility function clearly grows with the penetration level, whereas the empirical CDF remains relatively stable.

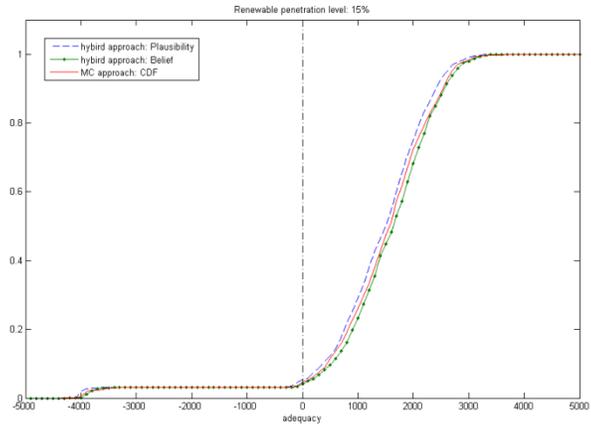

Fig 6. Comparison of joint propagation and pure probabilistic approaches at renewable penetration level of 15%

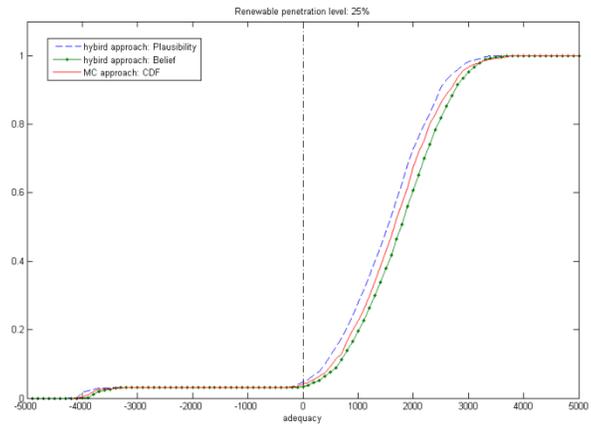

Fig 7. Comparison of joint propagation and pure probabilistic approaches at renewable penetration level of 25%

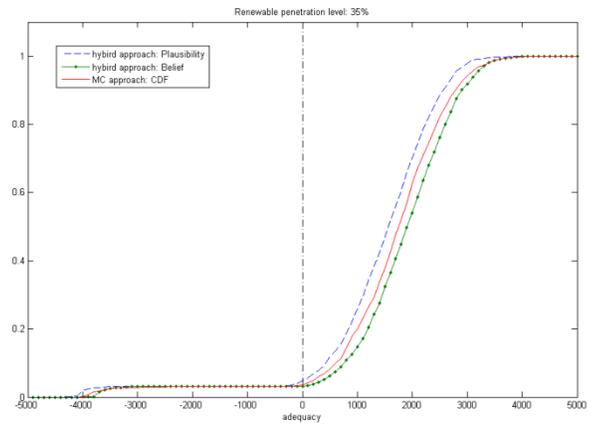

Fig 8. Comparison of joint propagation and pure probabilistic approaches at renewable penetration level of 35%

Table 4 summarizes the DG system unavailability values obtained by the hybrid and pure probabilistic methods under different renewable penetration levels. The unavailability is defined as the probability of the system adequacy being less than zero, and it is linearly correlated to other important adequacy assessment indices such as loss of load expectation (LOLE) and loss of energy expectation (LOEE) (Mabel et al. 2010). Two observations can be made: 1) the unavailability values from both methods decrease as the penetration level increases; 2) the separation between plausibility and belief, due to the epistemic uncertainty on the renewable generators parameters, grows as the penetration level increases. The former observation can be explained by the fact that the impact of transformer failure is reduced. The latter observation confirms that if the decision maker is interested in reducing the imprecision on the estimation of the system adequacy, he/she should try to improve the knowledge on the parameters of the renewable generators in the DG system.

Table 4. Comparisons of DG system unavailability at different renewable penetration levels

| Penetration levels | *Hybrid* | | *Pure probabilistic* |
|---|---|---|---|
| | *Plausibility* | *Belief* | |
| 15% | 0.0537 | 0.0408 | 0.0480 |
| 25% | 0.0482 | 0.0329 | 0.0410 |
| 35% | 0.0472 | 0.0315 | 0.0350 |

## 7. Conclusion and Discussion

This paper is a first attempt to develop a framework for thoroughly analyzing and treating the uncertainties in the adequacy assessment model of DG systems, to assist the DNOs' decision making processes. Two types of uncertainties (i.e. aleatory and epistemic) have been identified in the sample DG system considered and their representation has been described. Then, the joint propagation of the different representations of uncertainties (probabilistic and possibilistic) has been illustrated within the frame of evidence theory, which integrates the results in the form of plausibility and belief functions.

A numerical case study has been used to demonstrate the effects of the joint propagation, also in comparison with the pure probabilistic approach. As expected, the cumulative distribution of the DG system model output obtained by the pure probabilistic method lies within the belief and plausibility functions obtained by the joint propagation approach. Also, the imprecision in the DG parameters is explicitly reflected by the gap between the belief and plausibility functions. In addition, different levels of renewable penetrations have been considered, showing that the joint propagation approach captures well the growth of uncertainty when more DG resources penetrate the system, whereas the purely probabilistic empirical CDF remains stable.

These results imply that incorporating the imprecision existing in the definition of the parameters of the mathematical model due to incomplete knowledge, can be relevant for the DNOs' concerns in the management of distributed generation systems and can help substantiating his or her decision making.

As motivation for future research, we point out some of the main limitations of the study: 1) the joint propagation is developed by assuming independence among the probabilistic and possibilistic variables, and the independence within the probabilistic variables set; 2) dependence is introduced by the joint propagation algorithm among the possibilistic variables, because the same confidence level in the individual possibilistic variables is used to build the α–cuts; 3) the pure possibilistic model is not considered as terms of comparison, whereas it could be useful in the early stages of DG system design when there is very limited information available about the system characteristics.

**References**

AbdulHadi, M., Al-Ibrahim, A.M., Virk, G.S. 2004. Neuro-fuzzy-based solar cell model, IEEE Transactions on Energy Conversion, 19(3), 619 – 624.

Ackermann T., Andersson G., Söder L. 2001. Distributed generation: a definition. Electric Power Systems Research, 57(3), 195-204.

Apostolakis G. E. 1990. The concept of probability in safety assessments of technological systems. Science, 250(4986), 1359–1364.

Atwa, Y. M., El-Saadany, E. F., Salama, M. M. A., Seethapathy, R. 2010 Optimal Renewable Resources Mix for Distribution System Energy Loss Minimization. IEEE Transactions on Power Systems, 25(1), 360-370.

Aven T., Zio E. 2011a. Some considerations on the treatment of uncertainties in risk assessment for practical decision making Reliability Engineering & System Safety, 96(1), 64-74.

Aven T., Zio E. 2011b. Uncertainties in smart grids behavior and modeling: what risks and vulnerabilities? How to analyze them? Energy Policy, 39(10), 6308-6320.

Azbe V., Mihalic R. 2006. Distributed generation from renewable sources in an isolated DC network. Renewable Energy, 31(14), 2370-2384

Baraldi P., Zio E. 2008. A combined Monte Carlo and possibilistic approach to uncertainty propagation in event tree analysis. Risk Analysis, 28(5), 1309–25.

Baudrit, C., Dubois, D., Guyonnet, D. 2006. Joint propagation of probabilistic and possibilistic information in risk assessment. IEEE Transactions on Fuzzy Systems, 14, 593–608.

Billinton, R., Allan, R. N. 1996. Reliability Evaluation of Power Systems (2$^{nd}$ Edition). New York: Plenum.

Billinton, R., Gao, Y., Karki, R. 2009 Composite System Adequacy Assessment Incorporating Large-Scale Wind Energy Conversion Systems Considering Wind Speed Correlation. IEEE Transactions on Power Systems 24 (3) 1375-1382.

Cai Y.P., Huang G.H., Tan Q., Yang Z.F. 2009. Planning of community-scale renewable energy management systems in a mixed stochastic and fuzzy environment, Renewable Energy, 34(7), 1833-1847.

Catrinu M. D., Nordgard D. E. 2011. Integrating risk analysis and multi-criteria decision support under uncertainty in electricity distribution system asset management. Reliability Engineering and System Safety, 96, 663–670.

Conti S., Raiti S. 2007. Probabilistic load flow using Monte Carlo techniques for distribution networks with photovoltaic generators. Solar Energy, 81, 1473–1481.

Ding, Y., Wang, P., Goel, L., Loh, P. C., Wu, Q. 2011. Long-Term Reserve Expansion of Power Systems With High Wind Power Penetration Using Universal Generating Function Methods. IEEE Transactions on Power Systems, 26(2), 766-774.

Dubois D. 2006. Possibility theory and statistical reasoning. Computational Statistics & Data Analysis, 51, 47–69.

El-Khattam, W., Hegazy, Y.G., Salama, M.M.A. 2006. Investigating distributed generation systems performance using Monte Carlo simulation. IEEE Transactions on Power Systems, 21(2), 524 - 532.

Flage, R, Aven, T, Zio, E. 2009 Alternative representations of uncertainty in system reliability and risk analysis – Review and discussion. In: Martorell, S, Guedes Soares, C & Barnett, J (eds) Safety, Reliability and Risk Analysis: Theory Methods and Applications: Proceedings of the European Safety and Reliability Conference 2008 (ESREL 2008) and 17th SRA-Europe, Valencia, Spain, 22-25 September 2008.

Flage R., Baraldi P., Zio E., Aven T. 2010. Possibility-probability transformation in comparing different approaches to the treatment of epistemic uncertainties in a fault tree analysis. in: B. Ale, I.A. Papazoglu, E. Zio (Eds.), Reliability, Risk and Safety - Proceedings of the European Safety and RELiability (ESREL) 2010 Conference, Rhodes, Greece, 5-9 September 2010, pp. 694-698,.

Giannakoudis G., Papadopoulos A. I., Seferlis P., Voutetakis S. 2010. Optimum design and operation under uncertainty of power systems using renewable energy sources and hydrogen storage. International Journal of Hydrogen Energy, 35, 872-891.

Huang D., Chen T., Wang M. J. 2001. A fuzzy set approach for event tree analysis. Fuzzy Sets and Systems, 118, 153–165.

Hegazy, Y.G., Salama, M.M.A., Chikhani, A.Y. 2003. Adequacy Assessment of Distributed Generation Systems Using Monte Carlo Simulation. IEEE Transactions on Power Systems, 18(1), 48-52.

Hong, Y. Y., Pen, K. L. 2010 Optimal VAR Planning Considering Intermittent Wind Power Using Markov Model and Quantum Evolutionary Algorithm. IEEE Transactions on Power Delivery 25 (4) 2987-2996.

Karki, R., Hu, P., Billinton, R. 2010 Reliability Evaluation Considering Wind and Hydro Power Coordination. IEEE Transactions on Power Systems 25 (2) 685-693.

Leite da Silva A. M., da Fonscca Manso L. A., de Oliveira Mello J. C., Billinton R. 2000. Pseudo-Chronological Simulation for Composite Reliability Analysis with Time Varying Loads. IEEE Transactions on Power Systems 15 (1), 73-80.

Li Y.F., Zio E. 2011. A Multi-State Power Model for Adequacy Assessment of Distributed Generation via Universal Generating Function (submitted)

Mabel M. C., Raj R. E., Fernandez E. 2010. Adequacy evaluation of wind power generation systems. Energy, 35, 5217-5222.

Matos, M.A., Gouveia, E.M. 2008. The fuzzy power flow revisited. IEEE Transactions on Power Systems, 23(1), 213 – 218.

Massim, Y., Zeblah, A., Benguediab, M., Ghouraf, A., Meziane, R. 2006 Reliability evaluation of electrical power systems including multi-state considerations. Electrical Engineering 88 (2) 109-116.

Rose J., Hiskens I. A. 2008. Estimating wind turbine parameters and quantifying their effects on dynamic behavior. Proceedings of Power and Energy Society General Meeting. 1-7.

Saber A. Y., Venayagamoorthy G. K. 2011. Plug-in Vehicles and Renewable Energy Sources for Cost and Emission Reductions. IEEE Transactions on Industrial Electronics. 58(4), 1229 – 1238.

Shafer G., 1976. A Mathematical Theory of Evidence. Princeton, NJ: Princeton Univ. Press.

Soroudi A., Ehsan M. 2011. A possibilistic–probabilistic tool for evaluating the impact of stochastic renewable and controllable power generation on energy losses in distribution networks—A case study. Renewable and Sustainable Energy Reviews 15, 794–800.

Veliz, F.F.C., Borges, C.L.T., Rei, A.M. 2010. A Comparison of Load Models for Composite Reliability Evaluation by Nonsequential Monte Carlo Simulation. IEEE Transactions on Power Systems, 25(2), 649 - 656.

Zadeh, L. A. 1965. Fuzzy sets. Information and Control, 8, 338–353.

Zeng J., Liu J. F., Wu J., Ngan H.W. 2011. A multi-agent solution to energy management in hybrid renewable energy generation system. Renewable Energy, 36(5), 1352-1363.